\documentclass[aps,prd,twocolumn,nofootinbib,superscriptaddress]{revtex4-2}

\usepackage{amsfonts}
\usepackage{mathrsfs}
\usepackage{graphicx}
\usepackage{amsmath}
\usepackage{amssymb}
\usepackage{color}

\allowdisplaybreaks[4]

\usepackage[colorlinks=true,linkcolor=blue,citecolor=blue, urlcolor=blue]{hyperref}

\begin{document}

\title{Reanalysis of the top-quark pair production via the $e^+ e^-$ annihilation near the threshold region up to N$^3$LO QCD corrections}

\author{Jiang Yan}
\email{yjiang@cqu.edu.cn}
\author{Xing-Gang Wu}
\email{wuxg@cqu.edu.cn}
\author{Zhi-Fei Wu}
\email{wuzf@cqu.edu.cn}
\author{Jing-Hao Shan}
\email{sjh@cqu.edu.cn}
\address{Department of Physics, Chongqing Key Laboratory for Strongly Coupled Physics, Chongqing University, Chongqing 401331, P.R. China}

\author{Hua Zhou}
\email{zhouhua@cqu.edu.cn}
\address{School of Science, Southwest University of Science and Technology, Mianyang 621010, P.R. China}

\date{\today}

\begin{abstract}

In this paper, we present an improved analysis of the top-quark pair production via the process $e^{+}e^{-}\to \gamma^{*}\to t\bar{t}$ near the threshold region up to next-to-next-to-next-to-leading order (N$^3$LO) QCD corrections. Near the threshold region, the top-quark velocity $v$ tends to zero, leading to Coulomb singularity. To achieve a reasonable prediction in the threshold region, we reconstruct the analytical expression for the Coulomb-terms up to N$^{3}$LO accuracy by using the PSLQ algorithm, whose numerical values agree well with the previous N$^3$LO-level calculations. It is found that the N$^{3}$LO series still has sizable renormalization scale dependence, and to improve the precision of the series, we apply the Principle of Maximum Conformality to eliminate such scale dependence. After that, the Coulomb part is resummed into a Sommerfeld-Gamow-Sakharov factor, which finally leads to a much more reasonable behavior near the threshold region.

\end{abstract}

\maketitle

\section{Introduction}

The production of top-quark pair in electron-positron annihilation~\cite{Jersak:1981sp, Beenakker:1991ca, Chetyrkin:1996cf, Gao:2014nva, Gao:2014eea, Chen:2016zbz, Capatti:2022tit, Kiyo:2009gb, Beneke:2015kwa, Beneke:2016kkb,Beneke:2013jia, Czarnecki:1997vz, Hoang:1997ui, Lee:2018nxa, Fael:2022rgm, Fael:2022miw, Fadin:1987wz, Fadin:1988fn} holds substantial relevance for both the validation of the Standard Model (SM)~\cite{Schwienhorst:2022yqu} and the pursuit of new physics beyond it. As the heaviest species within the SM, a comprehensive probe into the diverse properties of the top quark is anticipated via the exploration of its threshold production. For instance, pinpoint theoretical estimations of the threshold production cross section for $e^{+}e^{-}\to t\bar{t}$ facilitate the ascertainment of the top quark's exact mass, which also serves as a vital observable for future high-energy electron-positron colliders such as the International Linear Collider (ILC)~\cite{ILC:2013jhg}, the Circular Electron Positron Collider (CEPC)~\cite{CEPCStudyGroup:2018ghi} and the Future Circular Collider~\cite{FCC:2018byv}, and etc.

Recently, the full next-to-next-to-next-to-leading order (N$^{3}$LO) QCD corrections to the heavy quark pair production $e^{+}e^{-}\to\gamma^*\to Q\bar{Q}$ has been completed in Ref.\cite{Chen:2022vzo}, which still exhibits significant renormalization scale dependence. Especially, the Coulomb-like interaction between the quark and the anti-quark triggers a pronounced enhancement of the cross section close to the production threshold, which has been incorporated to all orders in perturbation theory. To refine the current theoretical calculation, it is essential to address the mentioned renormalization scale dependence and the Coulomb singularity near the threshold. This can be accomplished by investigating renormalization scale-setting problem, estimating unknown higher-order (UHO) contributions, and resumming all of the Coulomb corrections. Such an endeavor will not only lead to a more robust theoretical calculation but also provide a more precise perturbative QCD (pQCD) prediction on the top-quark pair production cross section near the threshold region.

The fixed-order pQCD predictions for observables are usually assumed to suffer from an uncertainty in fixing the renormalization scale. This uncertainty in making fixed-order predictions occurs because one usually assumes an arbitrary renormalization scale together with an arbitrary range to ascertain its uncertainty. This simple treatment of the pQCD series and the renormalization scale causes the coefficients of the QCD running coupling ($\alpha_s$) at each order to be strongly dependent on the choice of both the renormalization scale and the renormalization scheme. The $\alpha_s$-running behavior is governed by the renormalization group equation (RGE),
\begin{align}
\beta(\alpha_{s})=\frac{{\rm d}\alpha_{s}(\mu^{2})}{{\rm d} \ln\mu^{2}} = -\sum_{i=0}^{\infty}\beta_{i}\alpha_{s}^{i+2}(\mu^{2}),
\end{align}
where the $\{\beta_{i}\}$-functions have been computed up to five-loop level in modified minimal-subtraction scheme~($\overline{\rm MS}$-scheme)~\cite{Gross:1973ju, Politzer:1974fr, Caswell:1974gg, Tarasov:1980au, Larin:1993tp, vanRitbergen:1997va, Chetyrkin:2004mf, Czakon:2004bu, Baikov:2016tgj, Herzog:2017ohr}. Consequently, the $\{\beta_{i}\}$-terms emerged in pQCD series can be reabsorbed into $\alpha_{s}$ to accurately fix the correct magnitude of $\alpha_s$. In respect of this fact, the Principle of Maximum Conformality~(PMC)~\cite{Brodsky:2011ta, Brodsky:2011ig, Mojaza:2012mf, Brodsky:2012rj, Brodsky:2013vpa} has been proposed to address the scale-setting problem in fixed-order pQCD calculation. It eliminates conventional renormalization scale ambiguity by using the above RGE recursively so as to set the correct $\alpha_s$ value and to achieve a well-matched (maximumly-approaching) conformal series between $\alpha_s$ and its expansion coefficients, yielding naturally scheme-independent theoretical predictions at any fixed-order, cf. the recent reviews~\cite{Wu:2013ei, Wu:2014iba, Wu:2019mky, DiGiustino:2023jiq}. This treatment is a kind of resummation of all the known type of RG-involved $\{\beta_i\}$-terms. More specifically, the PMC single-scale-setting approach (PMCs)~\cite{Shen:2017pdu, Yan:2022foz} determines a global effective coupling $\alpha_{s}(Q_{*}^{2})$~($Q_{*}$ is called as the PMC scale which can be viewed as the effective momentum flow of the process) for any fixed-order prediction by using the RGE, which also provides a solid way to extend the well-known Brodsky-Lepage-Mackenzie (BLM) approach~\cite{Brodsky:1982gc} to all orders~\footnote{A short review of the development of PMC from BLM and replies to some wrong comments on the PMC can be found in Ref.\cite{Brodsky:2023iap}.}. Using the PMCs, it has been demonstrated that the PMC prediction is independent to any choice of renormalization scheme and scale~\cite{Wu:2018cmb}, being consistent with the fundamental renormalization group approaches~\cite{Stueckelberg:1953dz, Peterman:1978tb, GellMann:1954fq, rge4} and also the self-consistency requirements of the renormalization group such as reflectivity, symmetry and transitivity~\cite{Brodsky:2012ms}.

It has been found that Coulomb-type corrections for the heavy-quark pair ($Q\bar{Q}$) production via the electron-positron annihilation have sizable contributions in the threshold region~\cite{Hagiwara:2008df, Kiyo:2008bv}, which similar to the hadronic production case~\cite{Brodsky:2012sz} shall also be enhanced by factors of $\pi$ and the resultant PMC scale can be relatively small when the heavy quark velocity $v\to 0$~\cite{Wang:2020ckr}. Thus the terms which are proportional to various powers of $(\pi/v)$ need to be treated carefully. In Ref.\cite{Wang:2020ckr}, the total cross section of $e^{+}e^{-}\to \gamma^{*}\to Q\bar{Q}$ near the threshold region has been investigated up to next-to-next-to leading order (N$^{2}$LO) level. At the N$^3$LO level and higher, the question will be much more involved due to the much more complex Coulomb structures~\cite{Beneke:2015kwa, Beneke:2016kkb, Beneke:2013jia}. In this paper, we will begin by partitioning the N$^{3}$LO-level total cross section of $e^{+}e^{-}\to \gamma^{*}\to Q\bar{Q}$ into two components, Coulomb and non-Coulomb corrections. Subsequently, we will employ the PMCs approach to deal with each components individually. Following that, we will perform a resummation of the divergent Coulomb corrections using the well-established Sommerfeld-Gamow-Sakharov~(SGS) factor~\cite{Sakharov:1948plh, Arbuzov:2011ff, Hoang:1997ui, Hoang:1997sj}. Then as an explicit example, we will give the numerical results for the case of top-quark pair production. Finally, we will utilize the Bayesian analysis (BA)~\cite{Cacciari:2011ze, Bagnaschi:2014wea, Bonvini:2020xeo, Duhr:2021mfd} to estimate the contributions from the UHO-terms.

The remaining parts of paper are organized as follows. In Sec.\ref{notion}, we present the total cross section of $e^{+}e^{-}\to \gamma^{*}\to Q\bar{Q}$ near the threshold region up to N$^{3}$LO-level, and explain how to deal with its Coulomb and non-Coulomb components separately. And then we derive the analytical expression of the Coulomb part by using the integer relation finding algorithm, e.g. the so-called PSLQ algorithm~\cite{1992A, 1999Analysis}. The results for the N$^{3}$LO-level Coulomb coefficients will be given in the Appendix. In Sec.\ref{pmc}, we give a concise introduction to the PMCs method. Numerical results and discussions for the top-quark pair production are given in Sec.\ref{results}. Sec.\ref{Summary} is reserved for a summary.

\section{The heavy quark pair production via the $e^+ e^-$ annihilation near the threshold region}\label{notion}

Up to N$^{3}$LO-level QCD corrections, the total cross section of $e^{+}e^{-}\to\gamma^{*}\to Q\bar{Q}$ can be written as
\begin{align}\label{Convseries}
	\sigma(s)=\sigma_{0}\left(1+r_{1}\alpha_{s}+r_{2}\alpha_{s}^{2}+r_{3}\alpha_{s}^{3} + {\cal O}(\alpha_s^4) \right),
\end{align}
where $\alpha_{s}\equiv\alpha_{s}^{(n_{l})}$ is QCD running coupling in $\overline{\rm MS}$-scheme with $n_{l}$ represents the active number of light flavors, and $\sigma_{0}$ is the LO cross section which takes the form
\begin{align}
	\sigma_{0}=N_{C} \frac{4\pi\alpha^{2}}{3s} \frac{v(3-v^{2})}{2} e_{Q}^{2}. \label{sigma0}
\end{align}
Here $N_{C}=3$ for $SU(3)_{C}$ color group, $\alpha$ is electromagnetic coupling constant, $v=\sqrt{1-4m_{Q}^{2}/s}$ is heavy quark velocity, $m_{Q}$ is the on-shell mass of the heavy quark $Q$, $s$ is the squared $e^+ e^-$ collision energy, and $e_{Q}$ is the charge of the heavy quark $Q$ in units of the electric charge, which equals to $+2/3$ for $Q=t$. It is convenient to express the perturbative coefficients $r_{1}$, $r_{2}$ and $r_{3}$ as
\begin{align}
	r_{1}&=\frac{1}{v}r_{1,v}+r_{1,+}, \label{r1}\\
	r_{2}&=\frac{1}{v^{2}}r_{2,v^{2}}+\frac{1}{v}r_{2,v}+r_{2,+}, \label{r2}\\
	r_{3}&=\frac{1}{v^{2}}r_{3,v^{2}}+\frac{1}{v}r_{3,v}+r_{3,+}, \label{r3}
\end{align}
where $r_{i,+}$ represents the terms of non-negative powers of $v$ in $r_{i}$ with $i=(1,2,3)$, respectively. Clearly, in the vicinity of the two-particle threshold, i.e., as the limit $v\to 0$, the total cross section exhibits the well-known Coulomb singularity, which is dominated by the negative power term of $v$ in Eqs.(\ref{r1},\ref{r2},\ref{r3}).

The analytical results of $r_{1}$, $r_{2}$ and all $n_{l}$-terms in $r_{3}$ can be found in Ref.\cite{Czarnecki:1997vz, Lee:2018nxa}, and the semi-numerical expression of $r_{3}$ can be found in Ref.\cite{Fael:2022rgm, Fael:2022miw}. In Ref.\cite{Chen:2022vzo}, numerical results up to N$^{3}$LO-level have been expressed as asymptotic expansions of $x=4m_{Q}^{2}/s$, which have been expanded up to $40_{\rm th}$-orders. Using those numerical results, we are able to reconstruct the analytical results of $r_{3}$ by using the PSLQ algorithm~\cite{1992A, 1999Analysis}. The PSLQ algorithm can be used to reconstruct analytical expression for given high-precision numerical values. As an explanation, the inputs to the PSLQ algorithm are high-precision numerical value denoted as $y$, and a vector $x = (x_{1}, \cdots, x_{n})$ consisting of real numbers that are analytically well-defined, such as $1$, $\ln 2$, $\pi$, ${\rm Li}_{n}(1/2)$, $\zeta_{n}$ and so on. Subsequently, the PSLQ algorithm determines a set of integer coefficients $m_{0}, m_{1}, \cdots, m_{n}$, such that the linear combination with these integer coefficients is close to zero within the required numerical precision, i.e., $m_{0}y+\sum_{i=1}^{n} m_{i}x_{i}=0$. Therefore, the analytical expression of $y$ can be written as $y=-\sum_{i=1}^{n} \frac{m_{i}}{m_{0}}x_{i}$. The PSLQ algorithm can be implemented using the Mathematica package \texttt{PolyLogTools}~\cite{Duhr:2019tlz}. Our results of those coefficients at the scale $\mu=\sqrt{s}$ are given in the Appendix.

Note that the coefficients in Eq.\eqref{Convseries} are parameterized in terms of $\alpha_{s}^{(n_{l})}$, which can be transformed from the results parameterized in terms of $\alpha_{s}^{(n_{l}+1)}$ by relating the coupling for the theory with $n_{l}+1$ flavors to the theory with $n_{l}$ flavors via the following decoupling relationship,
\begin{widetext}
	\begin{align}
		\frac{\alpha_{s}^{(n_{l}+1)}(\mu^{2})}{\alpha_{s}^{(n_{l})}(\mu^{2})} =1+\frac{1}{3}T_{F}\ln\frac{\mu^{2}}{m_{h}^{2}} \frac{\alpha_{s}^{(n_{l})}(\mu^{2})}{\pi}+T_{F}\left[-\frac{2}{9}C_{A} +\frac{15}{16}C_{F}+\left(\frac{5}{12}C_{A}+\frac{1}{4}C_{F}\right) \ln\frac{\mu^{2}}{m_{h}^{2}}+\frac{1}{9}T_{F}\ln^{2}\frac{\mu^{2}}{m_{h}^{2}}\right] \left(\frac{\alpha_{s}^{(n_{l})}(\mu^{2})}{\pi}\right)^{2},
	\end{align}
\end{widetext}
where $m_{h}$ is the on-shell mass of the decoupled heavy quark, $C_{A}=3$, $C_{F}=4/3$ and $T_{F}=1/2$.

The Coulomb singularity plays a dominant role for achieving precise total cross section near the threshold region ($v\to 0$). Following the suggestion of Refs.\cite{Czarnecki:1997vz, Hoang:1997ui, Hoang:1997sj}, we  schematically factorize the total cross section \eqref{Convseries} as the product of Coulomb and non-Coulomb parts, i.e.,
\begin{align}\label{factorization}
	\sigma=\sigma_{0} \times \mathcal{R}_{\rm C} \times \mathcal{R}_{\rm NC},
\end{align}
where $\mathcal{R}_{\rm C}$ and $\mathcal{R}_{\rm NC}$ represent the Coulomb part and the non-Coulomb part of the total cross section, respectively, which can be expressed as
\begin{align}
	\mathcal{R}_{\rm C}&=1+r_{1}^{\rm C}\alpha_{s}^{\rm V}(sv^{2})+r_{2}^{\rm C}\alpha_{s}^{\rm V,2}(sv^{2})+r_{3}^{\rm C}\alpha_{s}^{\rm V,3}(sv^{2}), \label{Coumseries}\\
	\mathcal{R}_{\rm NC}&=1+r_{1}^{\rm NC}\alpha_{s}(s)+r_{2}^{\rm NC}\alpha_{s}^{2}(s)+r_{3}^{\rm NC}\alpha_{s}^{3}(s).  \label{NCoumseries}
\end{align}
Here the QCD coupling $\alpha_{s}^{\rm V}(\vec{q}^{2})$ has been introduced for describing the interaction of the non-relativistic heavy quark-antiquark pair, which is defined as the effective charge in the following Coulomb-like potential~\cite{Appelquist:1977tw, Fischler:1977yf, Peter:1996ig, Schroder:1998vy, Smirnov:2008pn, Smirnov:2009fh, Anzai:2009tm, Kataev:2023sru}:
\begin{align}\label{V-potential}
	V(\vec{q}^{2})=-4\pi C_{F}\frac{\alpha_{s}^{\rm V}(\vec{q}^{2})}{\vec{q}^{2}},
\end{align}
where $\alpha_{s}^{\rm V}(\vec{q}^{2})$ absorbs all the higher-order QCD corrections, which is related to the $\overline{\rm MS}$-scheme coupling via the following way
\begin{align}
	\frac{\alpha_{s}^{\rm V}(\mu^{2})}{\alpha_{s}(\mu^{2})}=1+a_{1}\alpha_{s}(\mu^{2})+a_{2}\alpha_{s}^{2}(\mu^{2})+\mathcal{O}(\alpha_{s}^{3}),
\end{align}
where the expansion coefficients $a_{1}$ and $a_{2}$ have been calculated in Ref.\cite{Kataev:2023sru}, and their specific forms are as follows:
\begin{align}
	a_{1}=&\,\frac{1}{4\pi}\left(\frac{31}{9}C_{A}-\frac{20}{9}T_{F}n_{l}\right),\\
	a_{2}=&\,\frac{1}{(4\pi)^{2}}\bigg[\left(\frac{4343}{162}+4\pi^{2}-\frac{\pi^{2}}{4}+\frac{22}{3}\zeta_{3}\right)C_{A}^{2}\notag\\
	&\,-\left(\frac{1798}{81}+\frac{56}{3}\zeta_{3}\right)C_{A}T_{F}n_{l}-\left(\frac{55}{3}-16\zeta_{3}\right)C_{F}T_{F}n_{l}\notag\\
	&\,+\left(\frac{20}{9}T_{F}n_{l}\right)^{2}\bigg],
\end{align}
where $\zeta_{n}=\zeta(n)$ is Riemann's Zeta function. Then we can obtain
\begin{align}
	r_{1}^{\rm C}& = C_{F} \frac{\pi}{2v}, \label{r3C1}\\
	r_{2}^{\rm C}& = C_{F}^{2} \frac{\pi^{2}}{12 v^{2}}, \label{r3C2} \\
	r_{3}^{\rm C}& = C_{F} \left(\frac{\pi^{3}}{3 v} \beta_{0}^{2}(n_{l})-C_{F} \frac{2\zeta_{3}}{v^{2}} \beta_{0}(n_{l})\right), \label{r3C3}
\end{align}
and the coefficients $r_{i}^{\rm NC}$ can be obtained by expanding $\sigma(s)/(\sigma_{0}\mathcal{R}_{\rm C})$ over $\alpha_{s}$ and $v$.

\section{Simple introduction of principle of maximum conformality single-scale-setting method}\label{pmc}

In the previous literature~\cite{Shen:2017pdu, Yan:2022foz}, it has been found that the PMCs approach provides an all-orders single-scale-setting method to fix the conventional renormalization scale ambiguity, which adopts the standard RGE and the general QCD degeneracy relations among different orders~\cite{Bi:2015wea} to transform the RG-involved $n_f$-series into the $\{\beta_i\}$-series. In this subsection, we give a simple introduction of PMCs for self-consistency.

If the pQCD approximant $\rho(Q)$ starts and ends at the $n_{\rm th}$ and $(n+p-1)_{\rm th}$ order of $\alpha_{s}$, one has
\begin{align}
	\rho(Q)=\sum_{k=1}^{p} c_{k}\alpha_{s}^{n+k-1}(\mu_{R}^{2}),
\end{align}
where $\mu_{R}$ is an arbitrary initial choice of renormalization scale, $Q$ represents the kinetic scale at which the observable is measured or the typical momentum flow of the reaction. The perturbative coefficients $c_{k}$ can be rewritten as the following forms by using the general QCD degeneracy relations:
\begin{align}
	c_{1}&=c_{1,0},\label{deg-rel1}\\
	c_{2}&=c_{2,0}+n\beta_{0} c_{2,1},\label{deg-rel2}\\
	c_{3}&=c_{3,0}+(n+1)\beta_{0} c_{3,1}+\frac{n(n+1)}{2}\beta_{0}^{2} c_{3,2}+n\beta_{1} c_{2,1},\label{deg-rel3}\\
	&\cdots.\notag
\end{align}
Specifically, the fixed-order pQCD series $\rho$ can be further rewritten as the following compact form:
\begin{widetext}
\begin{align}\label{pQCD approximant}
\rho(Q)=\sum_{k=1}^{p}\hat{c}_{k,0} \alpha_{s}^{n+k-1}(\mu_{R}^{2}) +\sum_{k=1}^{p-1}\left[(n+k-1) \beta(\alpha_{s})\alpha_{s}^{n+k-2}(\mu_{R}^{2}) \sum_{i=1}^{p-k}(-1)^{i} \Delta_{n,k}^{(i-1)}\sum_{j=0}^{i} C_{i}^{j}\hat{c}_{k+i-j,i-j}L_{\mu_{R}}^{j}\right],
	\end{align}
\end{widetext}
where $L_{\mu_{R}}\equiv \ln(\mu_{R}^{2}/Q^{2})$, and the combinatorial coefficients are $C_{i}^{j}=j!/i!(j-i)!$. $\hat{c}_{i,0}=c_{i,0}$ is the conformal coefficients, $\hat{c}_{i,j\ge1}=c_{i,j}|_{\mu_{R}=Q}$ is non-conformal coefficients, $c_{i,j}$ and $\hat{c}_{i,j}$ can be connected as the relation $c_{i,j}=\sum_{k=0}^{j}C_{j}^{k} \hat{c}_{i-k,j-k}L_{\mu_{R}}^{k}$. $\Delta_{n,k}^{(i-1)}$ is related to the running of the RGE, which is defined as
\begin{align}
	\Delta_{n,k}^{(i-1)}(\mu_{R})=\frac{1}{(n+k-1)\beta\alpha_{s}^{n+k-2}}\frac{1}{i!}\frac{{\rm d}^{i}\alpha_{s}^{n+k-1}}{({\rm d}\ln \mu^{2})^{i}}\bigg|_{\mu=\mu_{R}},
\end{align}
we thus have
\begin{align}
	\Delta_{n,k}^{(0)}&=1,\\
	\Delta_{n,k}^{(1)}&=-\frac{1}{2!}\sum_{i=0}(n+k+i)\beta_{i}\alpha_{s}^{i+1},\\
	\Delta_{n,k}^{(2)}&=\frac{1}{3!}\sum_{j=0}\sum_{i=0}^{j}(n+k+i)(n+k+j+1)\beta_{i}\beta_{j-i}\alpha_{s}^{j+2},\\
	&\cdots.\notag
\end{align}

According to the PMCs procedure, all non-conformal terms in Eq.\eqref{pQCD approximant} should be absorbed into $\alpha_{s}(Q_{*}^{2})$, where $Q_{*}$ is a global effective scale and its value can be determined by requiring all non-conformal terms vanish in Eq.\eqref{pQCD approximant}, i.e., one can obtain PMC scale $Q_{*}$ by solving the following equation:
\begin{align}\label{PMCscale}
\sum_{k=1}^{\ell+1}&\,\bigg[(n+k-1)\alpha_{s}^{k-1}(Q_{*}^{2})\sum_{i=1}^{\ell-k+2}(-1)^{i}\Delta_{n,k}^{(i-1)}(Q_{*})\notag\\
	&\,\times\sum_{j=0}^{i}C_{i}^{j}\hat{c}_{k+i-j,i-j}L_{Q_{*}}^{j}\bigg]=0,
\end{align}
where $\ell$ indicates that the PMC scale $Q_{*}$ can be fixed up to N$^{\ell}$LL accuracy. Since the $\{\beta_{i}\}$-terms in the perturbative series (\ref{pQCD approximant}) have been resummed to determine the accurate value of $\alpha_s$, and the resultant PMC series is free of divergent renormalon terms and becomes scheme- and scale-independent pQCD conformal series~\cite{Wu:2018cmb}, e.g.
\begin{align}
	\rho|_{\rm PMC}=\sum_{k=1}^{p}\hat{c}_{k,0}\alpha_{s}^{n+k-1}(Q_{*}^{2}).
\end{align}
Practically, to achieve the analytic expression, the PMC scale $Q_{*}$ is expressed as the form of $\ln \left(Q_{*}^{2}/Q^{2}\right)$ by expanding Eq.\eqref{PMCscale} in terms of $\alpha_{s}(Q_{*}^{2})$ or $\alpha_{s}(Q^{2})$ up to the required N$^{\ell}$LL accuracy~\cite{Shen:2017pdu, Yan:2022foz}. In this paper, we will directly solve Eq.\eqref{PMCscale} numerically so as to derive more accurate PMC scales $Q_{*,\rm C}$ and $Q_{*,\rm NC}$~\footnote{Due to double suppression from both the $\alpha_s$-suppression and the exponential suppression, the usual analytic treatment is accurate enough. For the present case, the N$^{\ell}$LL-level expansion series of $\ln \left(Q_{*}^{2}/Q^{2}\right)$ does not convergent enough in the threshold region; It is thus helpful to keep all known divergent $1/v$-terms so as to achieve a more accurate prediction.}.

\section{Numerical Results}\label{results}

We are now ready to deal with the Coulomb and non-Coulomb corrections within the PMC framework.

For the Coulomb part $\mathcal{R}_{\rm C}$ (\ref{Coumseries}), after applying the PMC, we obtain the resultant PMC series, i.e.
\begin{eqnarray}
\mathcal{R}_{\rm C}|_{\rm PMC} &=& 1+\sum_{i} \hat{r}_{i,0}^{\rm C}\alpha_{s}^{{\rm V},i}(Q_{*,\rm C}^{2}),\notag\\
	&=& 1 + \frac{\pi}{2v} C_{F} \alpha_{s}^{\rm V}(Q_{*,\rm C}^{2}) + \frac{\pi^{2}}{12v^{2}} C_{F}^{2}
        \alpha_{s}^{\rm V,2}(Q_{*,\rm C}^{2}) \nonumber\\
    & & +{\cal O}(\alpha_{s}^{\rm V,4}),
\end{eqnarray}
whose first three conformal coefficients $\hat{r}_{i,0}^{\rm C}$ can be read from the coefficients $r_{i}^{\rm C}$ given in Eqs.(\ref{r3C1},\ref{r3C2},\ref{r3C3}) by applying the degeneracy relations (\ref{deg-rel1}-\ref{deg-rel3}) with the help of the replacements $c_{i}\to r_{i}^{\rm C}$. It is interesting to find that since $\hat{r}_{3,0}^{\rm C}= 0$, the N$^3$LO Coulomb coefficient $r_{3}^{\rm C}$~\eqref{r3C3} is exactly non-conformal, which can be resummed through the RGE and determines an effective $\alpha_s$, giving the required effective momentum flow near the threshold region (corresponding to the PMC scale $Q_{*,C}$). That is, the resultant PMC Coulomb series can be resummed as a Sommerfeld-Gamow-Sakharov (SGS) factor~\cite{Czarnecki:1997vz, Hoang:1997ui, Hoang:1997sj}~\footnote{The SGS factor~\eqref{PMC_c} can be expanded over $X$ as $1 +{X}/{2} +{X^{2}}/{12} -{X^{4}}/{720} + {\cal O}(X^5)$, in which the $X^{3}$-coefficient is exactly zero. As shown by Eq.(\ref{r3C3}), the coefficient $r_3^C$ is non-zero, thus the N$^3$LO-level Coulomb part of the series~\eqref{Coumseries} cannot be resummed into a simple SGS factor. Thus we will not provide the resummed Coulomb results for the predictions under conventional scale-setting approach.}:
\begin{equation}\label{PMC_c}
\mathcal{R}_{\rm C}|_{\rm PMC}^{\rm resum} = \frac{X}{1-e^{-X}}, \;\; X=\frac{\pi}{v} C_{F} \alpha_{s}^{\rm V}(Q_{*,\rm C}^{2}).
\end{equation}
The $1/v$ in the numerator can be canceled by the $v$ factor in the LO cross section $\sigma_{0}$, thus the Coulomb singularity near the threshold can be eliminated.

Similarly, for the non-Coulomb part (\ref{Coumseries}), its PMC conformal series up to N$^{3}$LO can be achieved and be written in the following form:
\begin{align}\label{PMC_nc}
	\mathcal{R}_{\rm NC}|_{\rm PMC}=1+\sum_{i=1}^{3}\hat{r}_{i,0}^{\rm NC}\alpha_{s}^{i}(Q_{*,\rm NC}),
\end{align}
where conformal coefficients $\hat{r}_{i,0}^{\rm NC}$ can be obtained in the same way as that of $\hat{r}_{i,0}^{\rm C}$.

It is worth mentioning that, even though the Coulomb divergence has been eliminated, $\ln v$-terms still remain in the $v^0$ terms of $\mathcal{R}_{\rm NC}$. However, when compared to the terms with negative powers of $v$, these terms are relatively more benign, and therefore, further treatment of those terms is not addressed in this article.

\subsection{The cross section of $t\bar{t}$ production}\label{ttbar}

To do numerical calculation, we take $m_{c}=1.5~{\rm GeV}$, $m_{b}=4.8~{\rm GeV}$, $m_{t}=172.69~{\rm GeV}$, $\alpha=1/132.2$ and $\alpha_{s}(M_{Z})=0.1179$ as their central values.

\begin{figure}[htbp]
\centering
\includegraphics[width=0.48\textwidth]{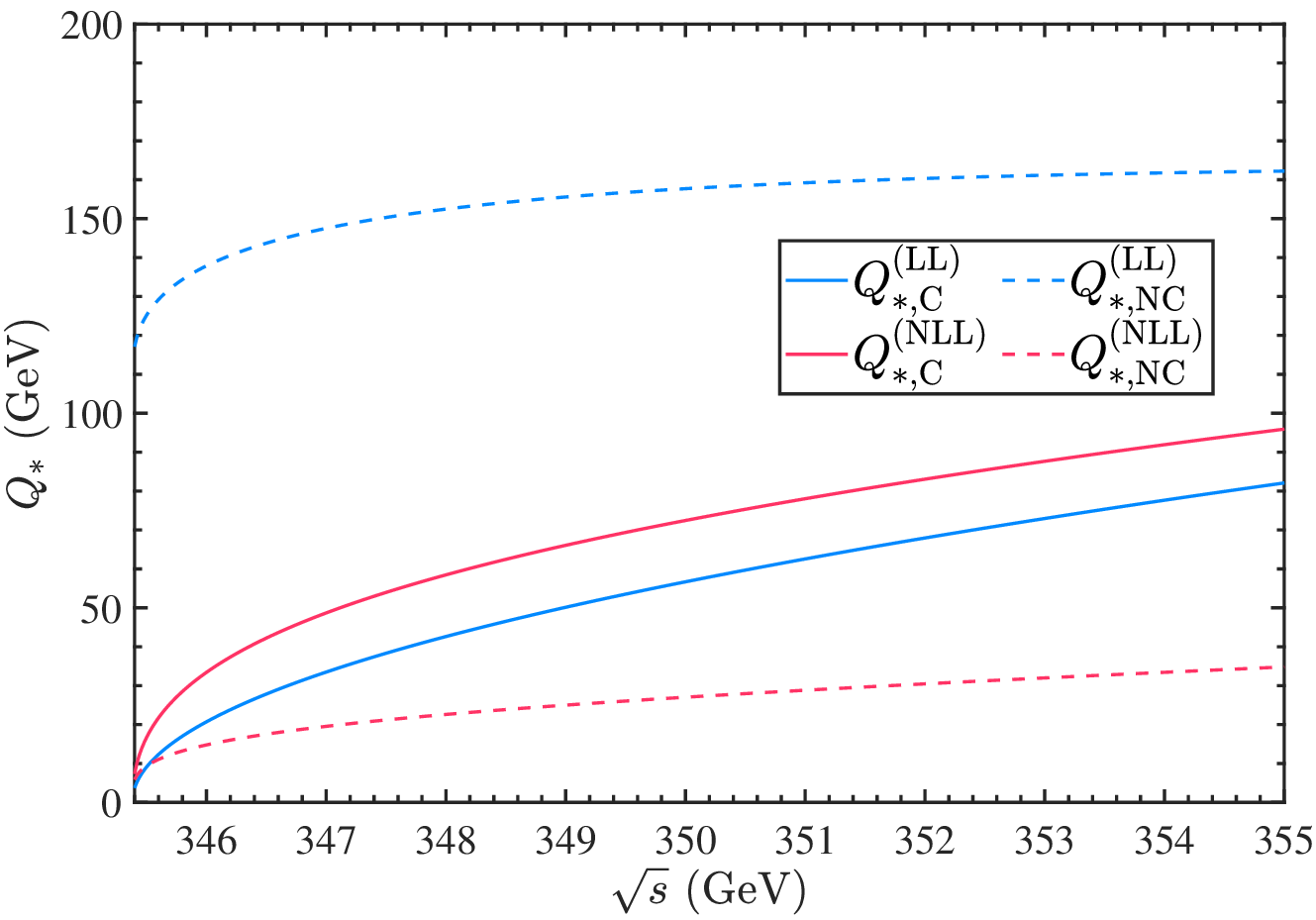}
\caption{(Color online) The PMC scales for the Coulomb and non-Coulomb corrections versus the $e^+ e^-$ collision energy $\sqrt{s}$. The blue dashed, the red dashed, the blue solid and the red solid lines represent the LL- and NLL- non-Coulomb and Coulomb scales, respectively.}
\label{Plot1}
\end{figure}

Following the procedures mentioned in previous section, we can numerically obtain the PMC scales versus the $e^+ e^-$ collision energy $\sqrt{s}$ for the Coulomb and non-Coulomb corrections, respectively. The results are presented in Fig.\ref{Plot1}. One may observe that the PMC scale $Q_{*,\rm C}$ for the Coulomb corrections is more sensitive to the changes in $\sqrt{s}$ (or $v$) than that of the non-Coulomb corrections, since both the LL-level and the NLL-level PMC scale $Q_{*,\rm C}$ are proportional to $\sqrt{s}v$.

\begin{figure}[htbp]
\centering
\includegraphics[width=0.48\textwidth]{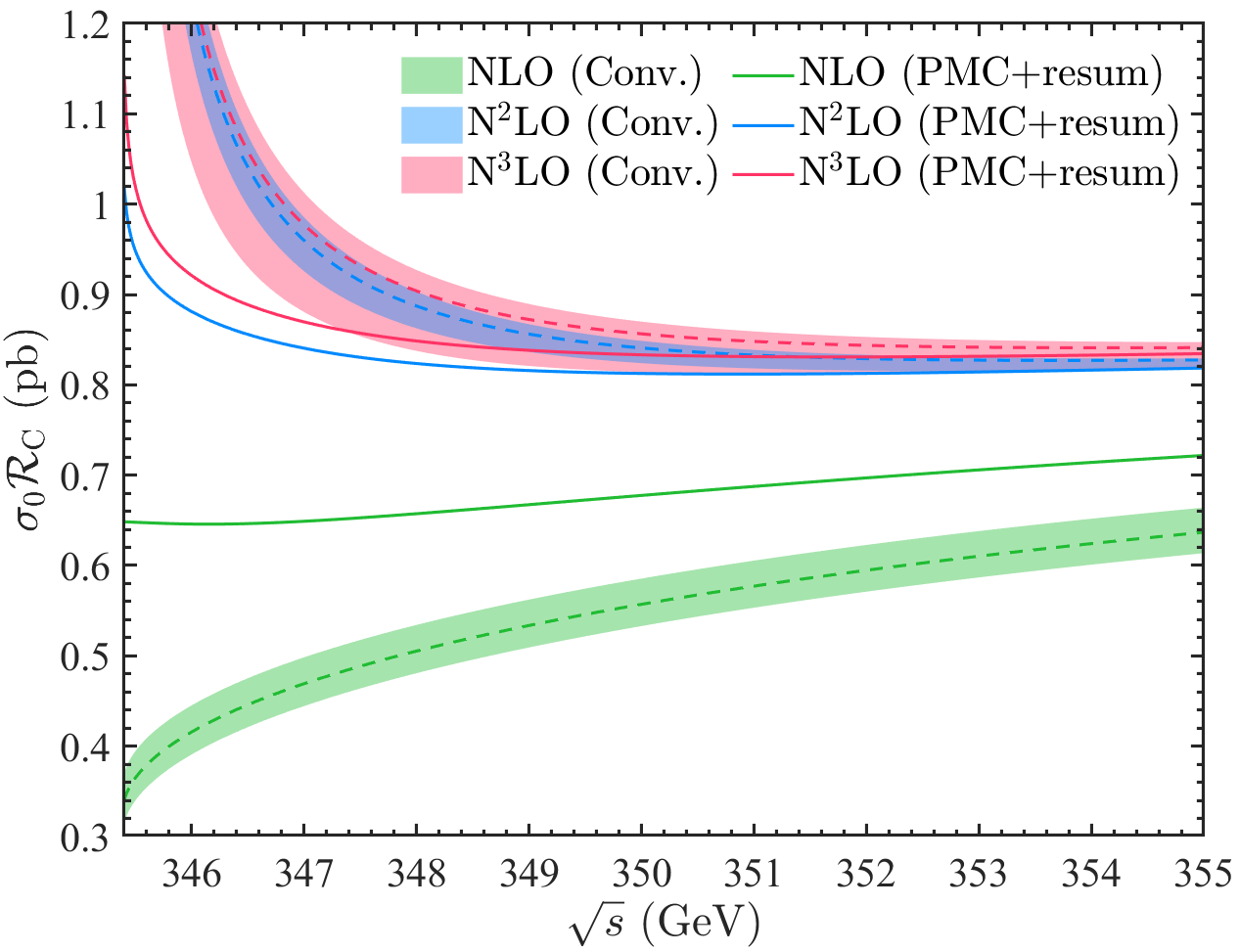}
\includegraphics[width=0.48\textwidth]{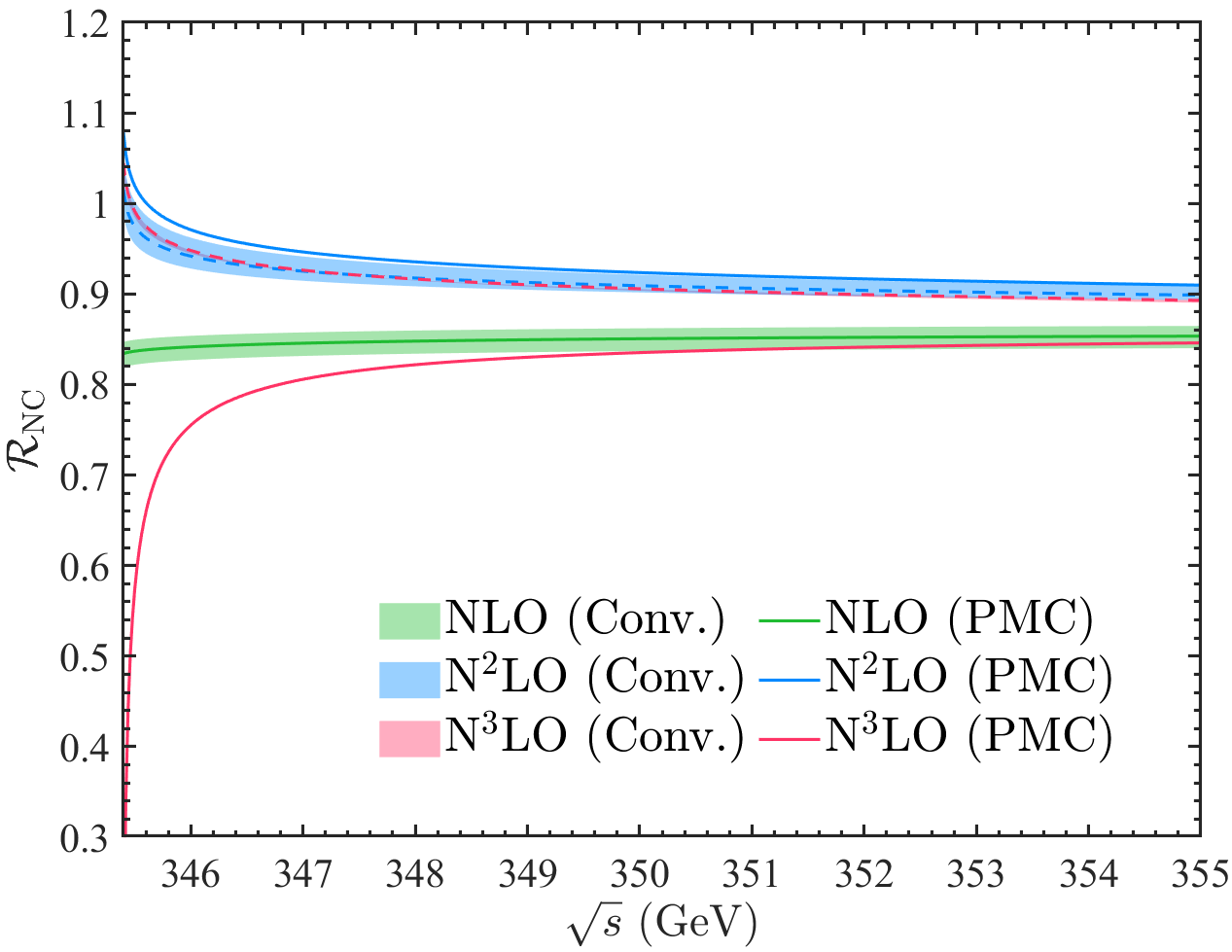}
\caption{(Color online) Comparisons of the Coulomb (upper) and the non-Coulomb (lower) contributions versus the $e^+ e^-$ collision energy $\sqrt{s}$ under conventional (Conv.) and PMC scale-setting approaches, respectively. The shaded bands for conventional series are for $\mu_{R}\in[\sqrt{s}/2, 2\sqrt{s}]$. }
\label{Plot4}
\end{figure}

To elucidate the behavior of the Coulomb and non-Coulomb contributions, we put them separately in Fig.\ref{Plot4}, where the results for both conventional series and PMC series are given. It is observed that to compare with the prediction of conventional series, by using the PMC scale-setting approach and resumming the divergent Coulomb terms, one will obtain a finite total cross section at the threshold point. And for the non-Coulomb sector, owing to the large cancellation between the conformal and the non-conformal terms, the scale-dependence and the logarithmic divergence are also mild for conventional series at the present N$^3$LO-level. It has been noted that there are still certain $\ln v$-terms in the non-Coulomb sector, and a more precise prediction by further resumming those divergent logarithmic terms may be achieved by applying the potential non-relativistic QCD~(pNRQCD)~\cite{Brambilla:2004jw, Brambilla:1999xf}.

\begin{figure}[htbp]
\centering
\includegraphics[width=0.48\textwidth]{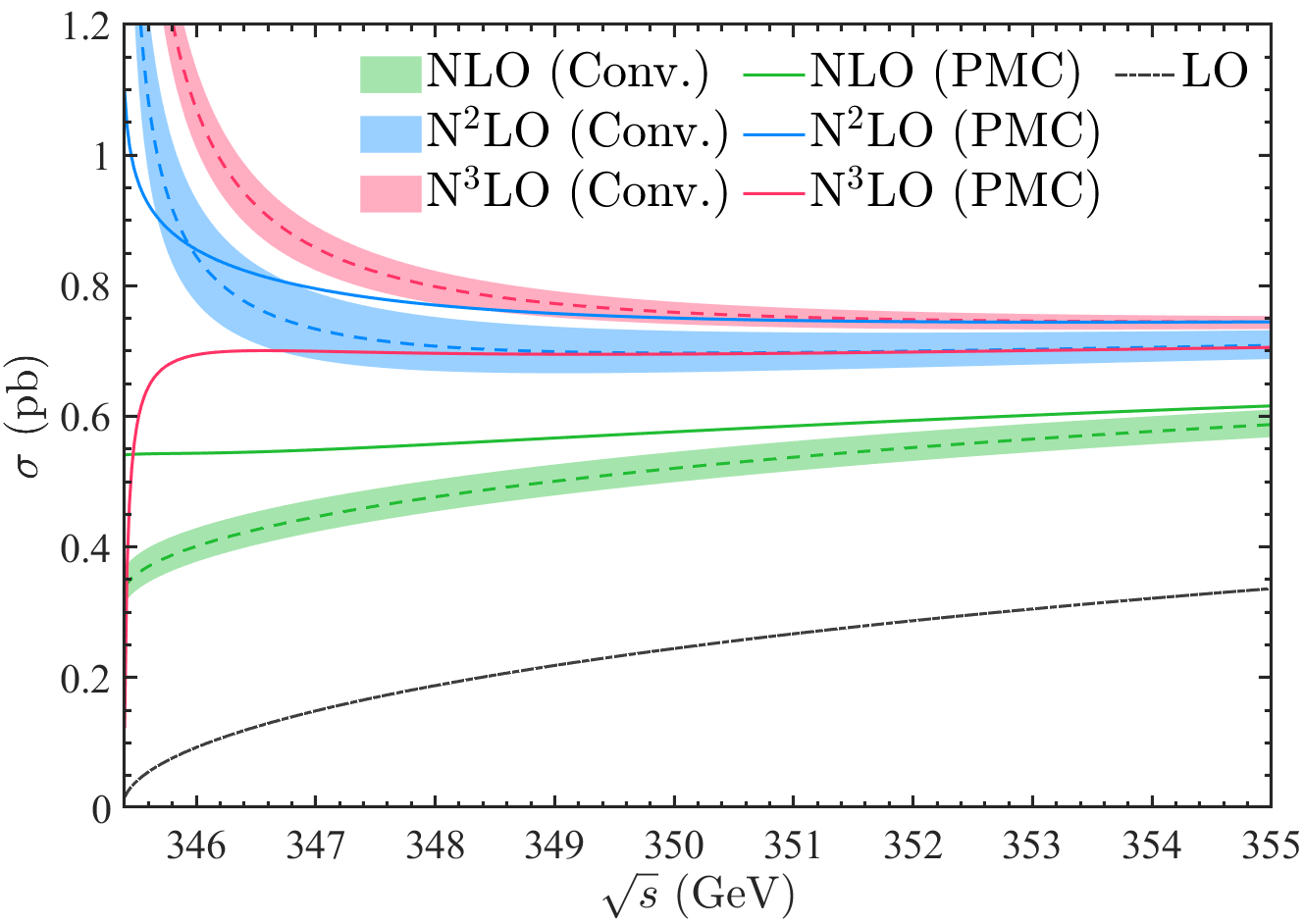}
\caption{(Color online) Total cross section $\sigma_{t\bar{t}}$ with QCD corrections up to N$^3$LO-level versus the $e^+ e^-$ collision energy $\sqrt{s}$ under conventional (dashed line) and PMC (solid line) scale-setting approaches, respectively. The shaded bands for conventional series are for $\mu_{R}\in[\sqrt{s}/2, 2\sqrt{s}]$.}
\label{Plot2}
\end{figure}

We then present the total cross section before and after applying the PMCs in Fig.\ref{Plot2}, where the middle lines correspond to the choice of $\mu_{R}=\sqrt{s}$ and the wide band correspond to the choice of $\mu_{R}\in[\sqrt{s}/2, 2\sqrt{s}]$ for conventional results. The application of PMC single-scale setting approach removes the renormalization scale dependence. By further doing the resummation of $1/v$-power terms, it shows that the divergency near the top-quark pair production threshold due to Coulomb-like interaction has been successfully suppressed.

\begin{figure}[htbp]
\centering
\includegraphics[width=0.48\textwidth]{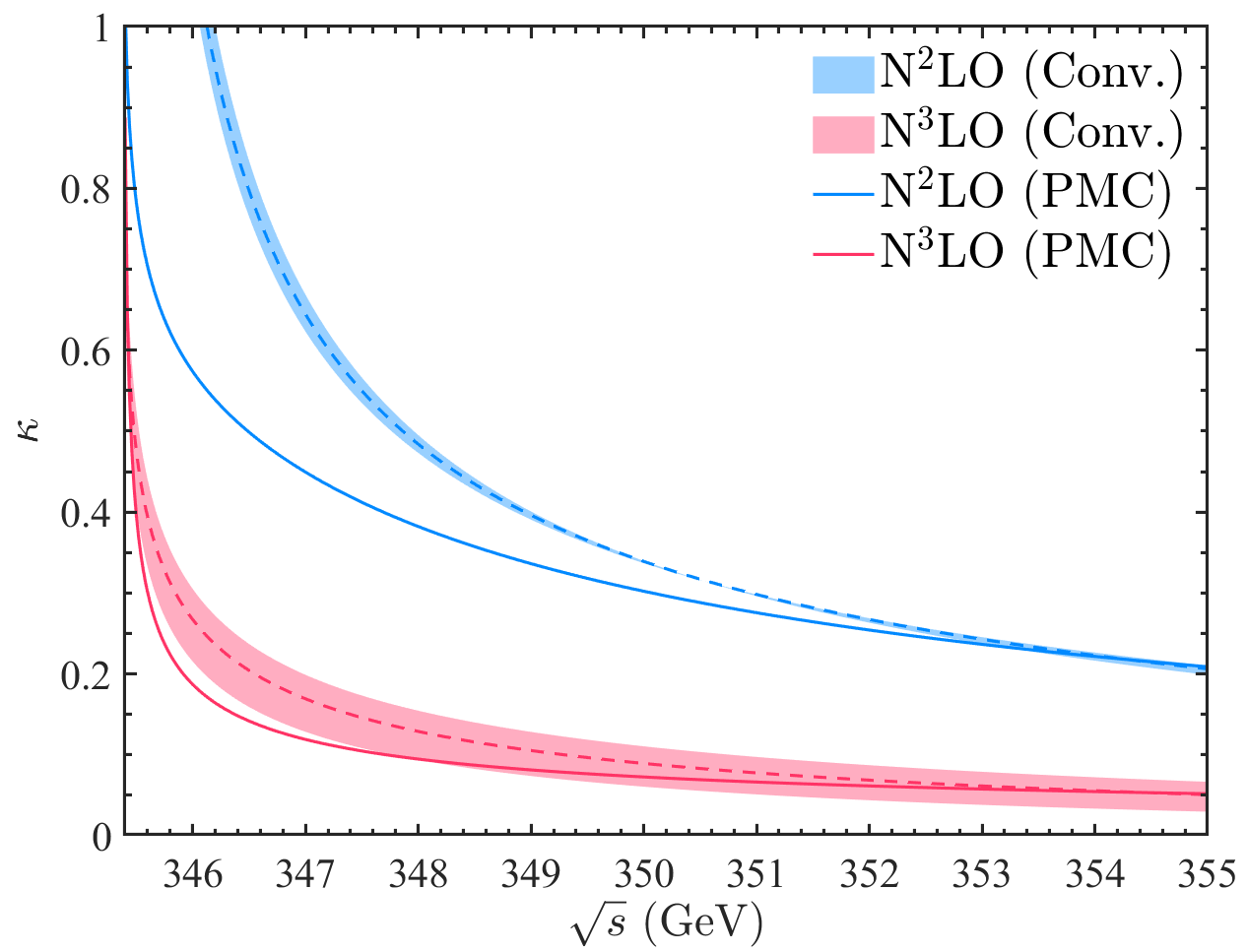}
\caption{(Color online) The factors $\kappa({\rm N^{2}LO})$ and $\kappa({\rm N^{3}LO})$ versus the $e^+ e^-$ collision energy $\sqrt{s}$ under conventional (dashed line) and PMCs (solid line) scale-setting approaches, respectively. The shaded bands for conventional series are for $\mu_{R}\in[\sqrt{s}/2, 2\sqrt{s}]$.}
\label{Plot5}
\end{figure}

For convenience, we define a $\kappa$-factor to characterize how the total cross section behaves when more loop terms have been taken into consideration, e.g.
\begin{equation}
\kappa({\rm N^{i}LO}) = \left|\frac{\sigma({\rm N^{i}LO})-\sigma({\rm N^{i-1}LO})}{\sigma({\rm N^{i-1}LO})}\right|.
\end{equation}
We present the results for $\kappa({\rm N^{2}LO})$ and $\kappa({\rm N^{3}LO})$ versus the $e^+ e^-$ collision energy $\sqrt{s}$ in Fig.(\ref{Plot5}). The $\kappa({\rm N^{3}LO})$ of conventional series strongly depends on the choice of renormalization scale $\mu_R$. After applying the PMC, the scale dependence is removed and the pQCD convergence near threshold region is improved.

\subsection{Estimation of the N$^4$LO QCD contributions using the Bayesian analysis}

The resulting PMC series offers not only precise predictions for the known fixed-order pQCD series but also provides a reliable foundation for assessing the possible contributions from the UHO-terms, thus greatly enhancing the predictive power of perturbation theory.

In the following, we will employ the Bayesian approach (BA)~\cite{Cacciari:2011ze, Bagnaschi:2014wea, Bonvini:2020xeo, Duhr:2021mfd} to estimate the magnitude of the uncalculated N$^{4}$LO-terms for the total cross section $\sigma(s)$ by utilizing the known N$^{3}$LO-level conventional and PMC series, respectively.

The BA quantifies the contributions of the UHO-terms in terms of the probability distribution. This method is powerful in constructing probability distributions, wherein Bayes' theorem is applied to iteratively update the probabilities as new information becomes available. A comprehensive introduction to the BA and its combination with the PMC approach can be found in Ref.\cite{Shen:2022nyr}. Generally, the conditional probability density function~(p.d.f.) $f_{r}(r_{p+1}|r_{1},\cdots,r_{p})$ for the uncalculated higher-order coefficient $r_{p+1}$ with known coefficients $\{r_{1},r_{2},\cdots,r_{p}\}$ is given by
\begin{align}
	f_{r}(r_{p+1}|r_{1},\cdots,r_{p})=\int h_{0}(r_{p+1}|\bar{r})f_{\bar{r}}(r_{p+1}|r_{1},\cdots,r_{p})\,{\rm d}\bar{r},
\end{align}
where $f_{\bar{r}}(r_{p+1}|r_{1},\cdots,r_{p})$ is the conditional p.d.f. for $\bar{r}$ with known coefficients $\{r_{1},r_{2},\cdots,r_{p}\}$ and $h_{0}(r_{p+1}|\bar{r})$ is the conditional p.d.f of $r_{p+1}$ given $\bar{r}$. Applying Bayes' theorem, we have
\begin{align}
	f_{\bar{r}}(r_{p+1}|r_{1},\cdots,r_{p})=\frac{h(r_{1},\cdots,r_{p}|\bar{r})g_{0}(\bar{r})}{\int h(r_{1},\cdots,r_{p}|\bar{r})g_{0}(\bar{r})\,{\rm d}\bar{r}},
\end{align}
where $h(r_{1},\cdots,r_{p}|\bar{r})=\prod_{k=1}^{p}h_{0}(r_{k}|\bar{r})$ and $g_{0}(\bar{r})$ are the likelihood function and the prior p.d.f. for $\bar{r}$ corresponding to the CH model~\cite{Cacciari:2011ze}, respectively. We can then calculate the degree-of-belief~(DoB) that the value of $r_{p+1}$ belongs to a certain credible interval~(CI) by using following formula
\begin{align}
	{\rm DoB} = \int_{-r_{p+1}^{(\rm DoB)}}^{r_{p+1}^{(\rm DoB)}}f_{r}(r_{p+1}|r_{1},\cdots,r_{p})\,{\rm d}r_{p+1}.
\end{align}
As a result, we obtain the expression for the boundary $r_{p+1}^{(\rm DoB)}$ of the symmetric smallest CI with fixed DoB for $r_{p+1}$ as
\begin{align}
	r_{p+1}^{(\rm DoB)}=\left\{
	\begin{aligned}
		&\bar{r}_{(p)}\frac{p+1}{p}{\rm DoB},&{\rm DoB}&\le \frac{p}{p+1}\\
		&\bar{r}_{(p)}\left[(p+1)(1-{\rm DoB})\right]^{-1/p},&{\rm DoB}&\ge \frac{p}{p+1}
	\end{aligned}\right.
\end{align}
where $\bar{r}_{(p)}={\rm max}\{|r_{1}|,|r_{2}|,\cdots,|r_{p}|\}$. Finally, the estimated $r_{p+1}$ for fixed DoB is
\begin{align}
	r_{p+1}\in\left[-r_{p+1}^{(\rm DoB)},r_{p+1}^{(\rm DoB)}\right].
\end{align}
In the subsequent calculation, ${\rm DoB}=95.5\%$ is taken to estimate the contributions from the UHO-terms.

\begin{figure}[htbp]
\centering
\includegraphics[width=0.48\textwidth]{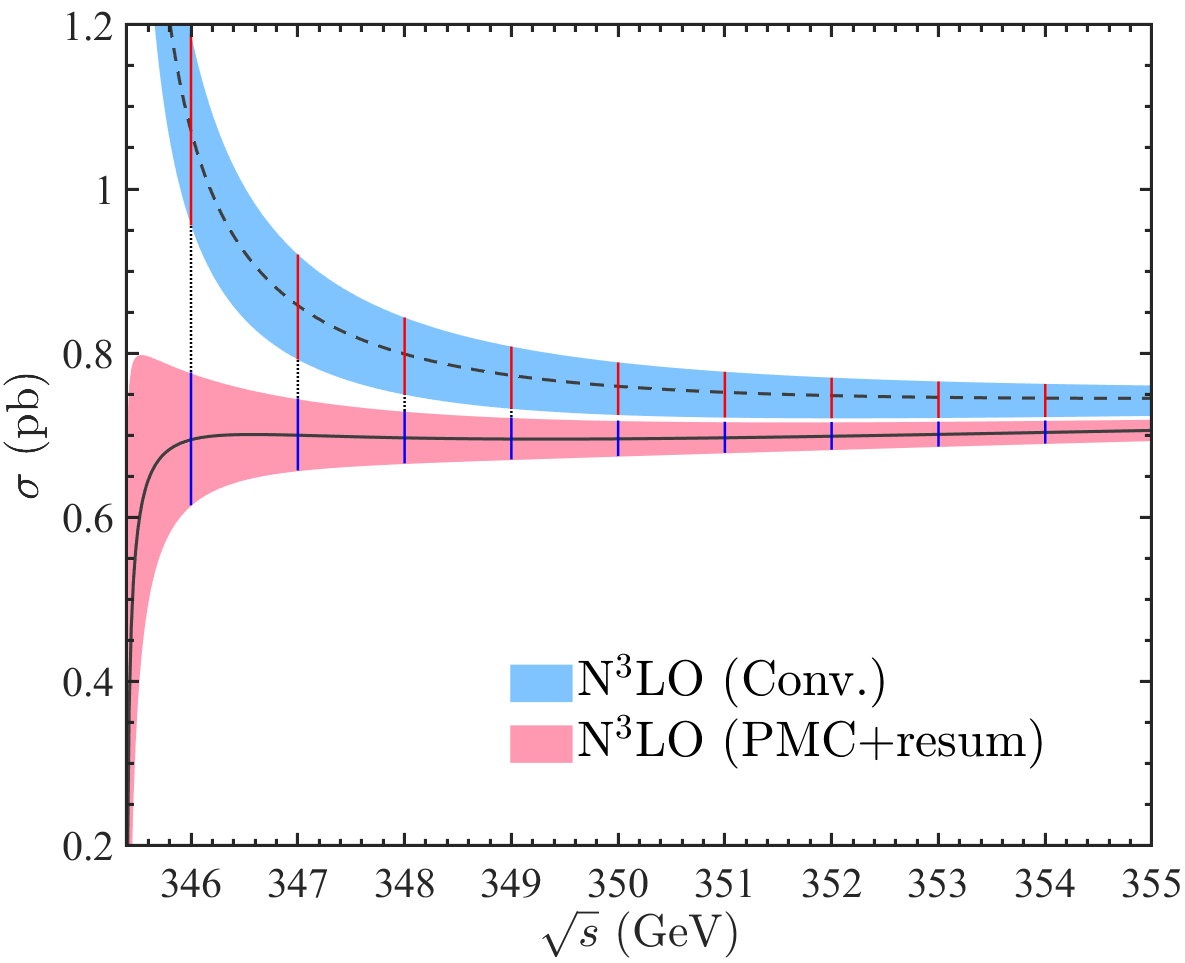}
\caption{The N$^3$LO-level total cross section $\sigma(s)$ versus the $e^+ e^-$ collision energy $\sqrt{s}$ under conventional (Dashed line) and PMCs (Solid line) scale-setting approaches, respectively. The vertical lines show magnitudes of the uncertainties caused by the UHO-terms.}
\label{Plot3}
\end{figure}

Since the Coulomb corrections have been resummed into a SGS factor, we only need to estimate the contributions from the UHO-terms for the non-Coulomb QCD series. Because the coefficients of the conventional pQCD series are known to be scale-dependent at every order, the BA can only be applied after one specifies the choices for the renormalization scale, thereby introducing extra uncertainties for the BA. Conversely, the PMC conformal series is scale-independent, providing a more reliable foundation for constraining predictions regarding the UHO contributions. By taking into account the conventional scale dependence for $\mu_{R}\in[\sqrt{s}/2, 2\sqrt{s}]$ and the predicted magnitudes of the UHO-terms, we present the uncertainties caused by the UHO-terms by using the BA in Fig.\ref{Plot3}. By applying the PMC, the uncertainties caused by the UHO-terms become smaller. These results confirm the importance of the PMC scale-setting approach.

\hspace{1cm}

\section{Summary} \label{Summary}

In the paper, we have presented an improved analysis for the total cross-section of $e^{+}e^{-}\to \gamma^{*}\to t\bar{t}$ near the threshold region up to N$^{3}$LO QCD corrections. Analytical expression for the N$^{3}$LO-level Coulomb-terms has been achieved by employing the PSLQ algorithm, which agrees well with the recently given numerical values in Ref.\cite{Chen:2022vzo}. The PMCs approach has been applied to eliminate the large renormalization scale uncertainties for the Coulomb and non-Coulomb terms separately. After that, the Coulomb-terms have been resummed as a Sommerfeld-Gamow-Sakharov factor, which greatly improves the reliability and precision of total cross-section near the threshold region.

Our results confirm the correctness and the importance of the PMC procedures for achieving scheme-and-scale invariant pQCD predictions, which can also be ensured by the commensurate scale relations among different schemes~\cite{Brodsky:1994eh, Huang:2020gic} and agrees with the renormalization group invariance~\cite{Petermann:1953wpa, peter2, Callan:1970yg, Symanzik:1970rt}. The elimination of the uncertainty in setting the renormalization scale for pQCD predictions using the PMCs, together with the reliable estimate for the UHO contributions obtained using the Bayesian analysis, greatly increases the precision of collider tests of the Standard Model and thus the sensitivity to new phenomena.

\hspace{2cm}

\noindent{\bf Acknowledgements}: This work was supported in part by the Chongqing Graduate Research and Innovation Foundation under Grant No.CYB23011 and No.ydstd1912, and by the Natural Science Foundation of China under Grant No.12175025 and No.12347101.

\section*{Appendix A}\label{A}

For the sake of completeness and convenience, we give the NLO and N$^{2}$LO coefficients $r_1$ and $r_2$ at the scale $\mu=\sqrt{s}$ in the following,
\begin{widetext}
	\begin{align}
		r_{1} = &\,\frac{C_{F}}{\pi} \bigg\{ \frac{1 + v^{2}}{v^{2}}\left[ 4 {\rm Li}_{2} \left(\frac{1 - v}{1 + v}\right) + 2 {\rm Li}_{2} \left( -\frac{1 - v}{1 + v}\right) + \ln\left( \frac{1 - v}{1 + v} \right) \left( \ln \left( \frac{2}{1 + v} \right) + 2\ln \left( \frac{2 v}{1 + v} \right) \right) \right]\notag\\
		&\qquad -2\left( \ln \left( \frac{2}{1 + v} \right) + 2 \ln \left( \frac{2 v}{1 + v} \right) \right) - \frac{(1 - v)(33 - 39v - 17 v^{2} + 7 v^{3})}{8 v (3 - v^{2})} \ln\left(\frac{1 - v}{1 + v}\right) + \frac{3 (5 - 3 v^{2})}{4 (3 - v^{2})} \bigg\}\\
		=&\,\frac{C_{F}}{\pi} \left[ \frac{\pi^{2}}{2 v} - 4 + \frac{\pi^{2}}{2} v + \left( - \frac{86}{9} + 8 \ln 2 + \frac{16}{3} \ln v \right) v^{2} + \mathcal{O}(v^{4}) \right],\\
		r_{2} = &\,C_{F} \bigg\{ \frac{\pi^{2} C_{F}}{12 v^{2}} + \frac{1}{v} \left[ - 2 C_{F} + C_{A} \left( \frac{31}{72} - \frac{11}{12} \ln v \right) + T_{F} n_{l} \left( - \frac{5}{18} + \frac{1}{3} \ln v \right) \right] - \left( C_{A} + \frac{2}{3} C_{F} \right) \ln v \notag\\
		&\qquad + C_{F}\left( - \frac{35}{18} + \frac{39}{4 \pi^{2}} + \frac{\pi^{2}}{6} + \frac{4}{3} \ln 2 - \frac{1}{\pi^{2}} \zeta_{3} \right) + C_{A} \left( \frac{179}{72} - \frac{151}{36\pi^{2}} - \left( \frac{8}{3} + \frac{22}{3\pi^{2}} \right) \ln 2 - \frac{13}{2 \pi^{2}} \zeta_{3} \right)\notag\\
		&\qquad + T_{F} n_{l} \left( \frac{11}{9\pi^{2}} +\frac{8}{3\pi^{2}} \ln 2\right) + T_{F} n_{h} \left( -\frac{4}{9} + \frac{44}{9\pi^{2}} \right)\notag\\
		&\qquad + \left[ \frac{839}{216} + \frac{173}{18} \ln 2 + \frac{101}{36} \ln v + T_{F} n_{l} \left( -\frac{13}{18} + \frac{1}{3} \ln v \right) + T_{F} n_{h} \left( \frac{1}{6} + \frac{1}{3}\ln2\right) \right] v + \mathcal{O}(v^{2}) \bigg\},
	\end{align}
\end{widetext}
where ${\rm Li}_{n}(z)=\sum_{k=1}^{\infty}z^{k}/k^{n}$ is polylogarithm function. As for the N$^3$LO coefficient $r_3$, the analytic expression of its coefficient $r_{3,+}$ with non-negative power terms of $v$ is very tedious and hard to give its analytic form, so we directly adopt its numerical form given by Ref.\cite{Chen:2022vzo}.

As for the N$^3$LO-level coefficients $r_{3,v}$ and $r_{3,v^{2}}$ with negative power terms $1/v$ and $1/v^2$ that are helpful for fixing the PMC scale of the Coulomb part and for getting the SGS factor, we adopt the PSLQ algorithm with the help of Mathematica package \texttt{PolyLogTools}~\cite{Duhr:2019tlz} so as to derive their analytic form, e.g.
\begin{widetext}
	\begin{align}
		r_{3,v^{2}}=&\, C_{F}\left[-\frac{\pi}{3}C_{F}^{2} + C_{F} C_{A} \left( \frac{31\pi}{216} - \frac{11}{6\pi} \zeta_{3} - \frac{11\pi}{36} \ln v \right) + C_{F}T_{F}n_{l} \left(-\frac{5\pi}{54} + \frac{2}{3\pi}\zeta_{3} + \frac{\pi}{9} \ln v \right) \right],\\
		r_{3,v}=&\, C_{F}\bigg\{ C_{F}^{2} \left(\frac{39}{8\pi} - \frac{35\pi}{36} +\pi \ln 2 - \frac{\zeta_{3}}{2\pi} - \frac{\pi}{3} \ln(2v)\right) + C_{A}^{2} \left(\frac{4343}{5184\pi} + \frac{175\pi}{432} - \frac{\pi^{3}}{128} + \frac{11}{48\pi}\zeta_{3} - \frac{247}{108\pi}\ln v + \frac{121}{72\pi} \ln^{2}v \right) \notag\\
		&\qquad + C_{A}C_{F}\left[-\frac{275}{72\pi} + \frac{179\pi}{144} - \left( \frac{4\pi}{3} + \frac{11}{3\pi} \right) \ln 2 - \frac{13}{4\pi} \zeta_{3} + \left( \frac{11}{3\pi} - \frac{\pi}{2} \right) \ln v \right] + C_{F}T_{F}n_{h} \left(\frac{22}{9\pi}-\frac{2\pi}{9}\right)\notag\\
		&\qquad + T_{F}n_{l} \left[ C_{F} \left( \frac{331}{288\pi} + \frac{\zeta_{3}}{2\pi} + \frac{4}{3\pi} \ln 2 - \frac{13}{12\pi} \ln v \right) + C_{A} \left( -\frac{899}{1296\pi} - \frac{11\pi}{54} - \frac{7\zeta_{3}}{12\pi} + \frac{217}{108\pi} \ln v - \frac{11}{9\pi} \ln^{2}v \right) \right]\notag\\
		&\qquad + T_{F}^{2}n_{l}^{2} \left[ \frac{25}{162\pi} + \frac{\pi}{27} - \frac{10}{27\pi} \ln v + \frac{2}{9\pi} \ln^{2}v\right] \bigg\}.
	\end{align}
\end{widetext}
It should be pointed out that the correctness of the PSLQ algorithm can be further tested by comparing the analytical results of $r_{1}$ and $r_{2}$ given by Ref.\cite{Czarnecki:1997vz, Lee:2018nxa} with the numerical results given in Ref.\cite{Chen:2022vzo}; and by comparing with the above $r_3$ with the semi-numerical expression of $r_{3}$ given in Refs.\cite{Fael:2022rgm, Fael:2022miw}.

\end{document}